\documentclass[11pt,titlepage,a4paper]{article}
\usepackage{bozhomac}  
\usepackage{bozhlogo}  
\usepackage{cite}	
\usepackage{varioref}	

\title{\bfseries	\vspace*{-1.678902345in}
{\huge A mathematical base for\\[1ex] Fibre bundle formulation of\\[1ex]
				Lagrangian quantum field theory}
}

\vspace{1.7ex}

\author{
Bozhidar Z.\ Iliev
\thanks{Laboratory of Mathematical Modeling in Physics,
Institute for Nuclear Research and \mbox{Nuclear} Energy,
Bulgarian Academy of Sciences,
Boul. Tzarigradsko chauss\'ee~72, 1784 Sofia, Bulgaria}
\thanks{E-mail address: bozho@inrne.bas.bg}
\thanks{URL: http://theo.inrne.bas.bg/$\sim$bozho/}
}

\date{	
 \vspace{2.27ex}\ShortTitle{Mathematics for Bundle Quantum Field
Theory}\\[0.27ex]
 \vspace{3.27ex}
\small
\normalsize
\vspace{0.27ex}
\textsl{\bfseries
Report presented at the 10$^\text{the}$ International Workshop on\\
``Complex Structures, Integrability and Vector Fields''\\
Sofia, Bulgaria, 13 -- 17 September, 2010}		\\[2ex] %
\small
	\begin{tabular}{r@{$\colon~$}l}
\normalsize\sffamily\bfseries
	\end{tabular} \\[-0.27ex]
\normalsize
 \vspace{4.27ex}{\Huge\BOZHO}	\\[4.27ex]
 \vspace{0.27ex}\Subject{Quantum field theory, Differential geometry}
								\\[2.27ex]
	\begin{tabular}{r@{\hspace{0.512em}}|@{\hspace{0.512em}}l}
 \vspace{0.27ex}\MSC[2000]{53C80, 53C99, 53Z05\\
			   55R99, 91T99, 81Q70 }
&
 \vspace{0.27ex}\PACS[2003]{02.40.Ma, 02.90.+p\\
			     03.70.+k, 11.10.-z}
	\end{tabular} \\[1.27ex]
 \vspace{0.27ex}\KeyWords{Quantum field theory, Heisenberg picture of motion\\
		Fibre bundles, Geometrization of quantum field theory}
								\\[0.27ex]
}

\newcommand{\Hil}{\mathcal{F}}		
\newcommand{\HilB}{(\bHil,\proj,\base,\Hil)}	
	\newcommand{\bHil}{\mathit{F}}	
	\newcommand{\proj}{\pi}		
	\newcommand{\base}{\mathit{M}}	

\newcommand{\dyn}[1]{\pmb{\mathbb{#1}}}	

\newcommand{\ope}[2][{}]{\lindex[\mathcal{#2}]{}{#1}} 
\newcommand{\mor}[2][{}]{\lindex[\mspace{-2.8mu}\mathit{#2}]{}{#1}}
\newcommand{\hmor}[2][{}]{\mor[#1]{\Hat{#2}}}	
\newcommand{\lcl}{\lindex[\mspace{-0.75mu}l]{}{\circ}}
\newcommand{\lcD}{\lindex[\mspace{-1.75mu}D]{}{\circ}}

\begin{document}		

\renewcommand{\thepage}{\roman{page}}

\renewcommand{\thefootnote}{\fnsymbol{footnote}} 
\maketitle				
\renewcommand{\thefootnote}{\arabic{footnote}}   

\tableofcontents		


	\begin{abstract}

	The paper contains a differential-geometric foundations for an attempt
to formulate Lagrangian (canonical) quantum field theory on fibre bundles. In
it the standard Hilbert space of quantum field theory is replace with a Hilbert
bundle; the former playing a role of a (typical) fibre of the letter one.
Suitable sections of that bundle replace the ordinary state vectors and the
operators on the system's Hilbert space are transformed into morphisms of the
same bundle. In particular, the field operators are mapped into corresponding
field morphisms.

	\end{abstract}

\renewcommand{\thepage}{\arabic{page}}

\section {Introduction}
\label{Introduction}

	The purpose of this work is to be presented grounds for a
consistent formulation of quantum field theory in terms of fibre bundles. The
 ideas for that goal are shared
from~\cite{
	bp-BQM-1,bp-BQM-2,
	bp-BQM-3,bp-BQM-4,
	bp-BQM-5,
	bp-BQM-full},
where the quantum mechanics is formulated
on the geometrical language of fibre bundle theory.

In section~\ref{Subsect2.1} contains some basic definitions. Special
attention is paid on the Hilbert bundles, which will replace the Hilbert spaces
of the ordinary quantum field theory, and the metric structure in them.  In
section~\ref{Subsect2.2} is considered an isomorphism
between the fibres of a Hilbert bundle, called the bundle transport. It will
play a central role in this investigation.
	Sect.~\ref{Sect3} contains a motivation why the (Hilbert) fibre
bundles are a natural scene for a mathematical formulation of quantum
field theory.
	Sect.~\ref{Conclusion} concludes the paper.

\vspace{1.4ex}



\section{Fibre Bundles. Hilbert bundles}
	\label{Subsect2.1}

	To begin with, we present some facts from the theory of fibre
bundles~\cite{Husemoller,Steenrod}, in particular the Hilbert ones which will
replace the Hilbert spaces in ordinary quantum field theory.

	A \emph{bundle} is a triple $(E,\pi,B)$ of sets $E$ and $B$, called
(total) bundle space and base (space) respectively, and (generally)
surjective mapping $\pi\colon E\to B$, called projection. If $b\in B$,
 $\pi^{-1}(b)$ is the fibre over $b$ and, if
	 $Q\subseteq B$,
$(E,\pi,B)|_Q := (\pi^{-1}(Q),\pi|_{\pi^{-1}(Q)},Q)$
is the restriction on $Q$ of a bundle $(E,\pi,B)$.
	A \emph{section} of $(E,\pi,B)$ is a mapping
$\sigma\colon B\to E$ such that $\pi\circ\sigma=\id_B$, where  $\id_Z$
is the identity mapping of a set $Z$, and their set is denoted by
$\Sec(E,\pi,B)$.
 The set of morphisms of $(E,\pi,B)$ is\[
\Morf(E,\pi,B)
 := \{
	(\varphi,f)| \varphi\colon E\to E,\ f\colon B\to B,\
	\pi\circ\varphi = f\circ\pi
    \} .
\]
The set
of all $B$-morphisms (strong morphisms) of $(E,\pi,B)$
is
\(
 \Morf_B(E,\pi,B)
 := \{\varphi|\varphi\colon E\to E,\ \pi\circ\varphi=\pi \} .
\)
  	Consider the set of \emph{point-restricted morphisms}
	\begin{align*}
E_0 : &= \{
 (\varphi_b,f)\,|\,
	\varphi_b = \varphi|_{\pi^{-1}(b)},
	\ b\in B,\ (\varphi,f)\in\Morf(E,\pi,B)
	\}
\\
 & = \{
	(\varphi_b,f) \,|\, \varphi_b\colon\pi^{-1}(b)\to \pi^{-1}(f(b)),
	\ b\in B,\ f\colon B\to B
	\} .
	\end{align*}
Defining  $\pi_0\colon E_0\to B$ by $\pi_0(\varphi_b,f):=b$ for
$(\varphi_b,f)\in E_0$, we see that $\morf(E,\pi,B):=(E_0,\pi_0,B)$ is a fibre
bundle. This is the \emph{bundle of
point\ndash restricted morphisms} of $(E,\pi,B)$.
\footnote{~%
	There exists a bijective correspondence $\rho$ such that
\(
\Morf(E,\pi,B)  \xrightarrow{\ \rho\ } \Sec\bigl(\morf(E,\pi,B)\bigr) .
\)%
}

 The \emph{bundle $\morf_B(E,\pi,B)$ of
point\ndash restricted morphisms over} $B$ of $(E,\pi,B)$ has a base $B$,
bundle space
	\begin{align*}
E_{0}^{B} : &=
 \{
   \varphi_b \,|\, \varphi_b=\varphi|_{\pi^{-1}(b)},\
	b\in B,\ \varphi\in\Morf_B(E,\pi,B)
 \}
\\
 & = \{ \varphi_b \,|\, \varphi_b\colon \pi^{-1}(b)\to\pi^{-1}(b),\ b\in B \}
	\end{align*}
and projection $\pi_{0}^{B}\colon E_{0}^{B}\to B$ such that
\(
\pi_{0}^{B}(\varphi_b) := b,\quad \varphi_b\in E_{0}^{B} .
\)
The bundle $\morf_B(E,\pi,B)$ will be refereed as the
\emph{bundle of restricted morphisms} of $(E,\pi,B)$.
There is a bijection
\(
\Morf_B(E,\pi,B)  \xrightarrow{\ \chi\ } \Sec\bigl(\morf_B(E,\pi,B)\bigr)
\)
given by
$\chi\colon\varphi\mapsto\chi_\varphi$, $\varphi\in\Morf_B(E,\pi,B)$,
with $\chi_\varphi\colon b\mapsto\chi_\varphi(b):=\varphi|_{\pi^{-1}(b)}$,
$b\in B$. Its inverse is
$\chi^{-1}\colon\sigma\mapsto\chi^{-1}(\sigma)=\varphi$,
$\sigma\in\Sec(\morf_B(E,\pi,B))$, with $\varphi\colon E\to E$ given via
$\varphi|_{\pi^{-1}(b)}=\sigma(b)$ for every $b\in B$.

	A mapping $\Sec(E,\pi,B)\to\Sec(E,\pi,B)$ will be called
\emph{morphism of the set $\Sec(E,\pi,B)$ of sections of bundle}
$(E,\pi,B)$. The set of all such mappings will be denoted by
$\MorfSec(E,\pi,B)$.

	There is a natural mapping from $\Morf_B(E,\pi,B)$
into $\MorfSec(E,\pi,B)$ given for $\mor{A}\in\Morf_B(E,\pi,B)$ and
$\mor{X}\in\Sec(E,\pi,B)$ by $\hmor{A}\in\MorfSec(E,\pi,B)$
by $\hmor{A}(\mor{X}):=\mor{A}\circ\mor{X}$.
We say that $\hmor{A}$ as a \emph{morphism of sections (of
$(E,\pi,B)$) induced or generated by $\mor{A}$}.

        	When \emph{vector bundles} are considered, in the definition of
a morphism or $B$-morphism is included the condition that the corresponding
fibre mappings must be \emph{linear}.
	\begin{Defn}	\label{Defn2.1}
	A \emph{Hilbert (fibre) bundle} is a vector bundle whose fibres over
the base are isomorphic Hilbert spaces or, equivalently, whose (standard)
fibre is a Hilbert space.
	\end{Defn}

       	Some quite general aspects of the Hilbert bundles can be found
in~\cite[chapter~VII]{Lang/Manifolds}.

	Let $\HilB$ be a Hilbert bundle with bundle space $\bHil$, base
$\base$, projection $\proj$, and (typical) fibre $\Hil$. The fibre over
$x\in\base$ will be often denoted by $\bHil_x$, $\bHil_x:=\proj^{-1}(x)$. Let
$l_x\colon\bHil_x\to\Hil$, $x\in\base$, be the isomorphisms defined by the
restricted decomposition functions, viz., as
$\phi_W|_x\colon\{x\}\times\Hil\to\bHil_x$ with $W$ being a neighborhood of
$x$, we define $l_x$ via
 $\phi_W|_x(x,\psi)=:l_x^{-1}(\psi)\in\proj^{-1}(x)$ for every $\psi\in\Hil$
and call $l_x$
\emph{point\ndash trivializing mappings (isomorphisms)}.

	Let $\clangle\,\cdot\, | \,\cdot\,\crangle\colon\Hil\times\Hil\to\mathbb{R}$ be
the scalar product in the Hilbert space
$\Hil$ and for every $x\in\base$ the mapping
\(
\langle\,\cdot\, | \,\cdot\,\rangle_x
	\colon\bHil_x\times\bHil_x\to\mathbb{R}
\)
be the scalar product in the fibre $\bHil_x$ considered as a Hilbert space.
 The vector structure of $\HilB$ is called \emph{compatible}
with its metric structure if the isomorphisms
$l_x\colon\bHil_x\to\Hil$ preserve the scalar products, \viz iff
\(
\langle\varphi_x | \psi_x\rangle_x
		= \clangle l_x(\varphi_x) | l_x(\psi_x) \crangle
\)
for every $\varphi_x,\psi_x\in\bHil_x$. A Hilbert bundle with compatible
vector and metric structure will be called \emph{compatible Hilbert bundle}.
In such a bundle the isomorphisms $l_x$, $x\in\base$ transform the
metric structure $\clangle \,\cdot\, | \,\cdot\, \crangle$ from $\Hil$ to
$\bHil$ and v.v.\ according to
	\begin{gather}	\label{2.1}
\langle \,\cdot\,|\,\cdot\, \rangle _x=
\clangle l_{x}\,\cdot\,|l_{x}\,\cdot\, \crangle, \qquad x\in \base
\\
	\tag{\ref{2.1}$^\prime$}	\label{2.1'}
\clangle \,\cdot\,|\,\cdot\, \crangle =
\langle l_{x}^{-1}\,\cdot\,|l_{x}^{-1}\,\cdot\, \rangle _x, \qquad x\in \base.
	\end{gather}
The mappings
$l_{x\to y}:=l_y^{-1}\circ l_x\colon\proj^{-1}(x)\to\proj^{-1}(y)$
are
(i) fibre mappings for fixed $y$,
(ii) linear isomorphisms, and
(iii) isometric, \ie metric preserving in a sense that
	\begin{equation}	\label{2.2}
\langle l_{x\to y} \,\cdot\, | l_{x\to y} \,\cdot\, \rangle_y
 = \langle\,\cdot\,  | \,\cdot\, \rangle_x.
	\end{equation}
  Consequently, all of the fibres over the base and the standard
fibre of a compatible Hilbert bundle are (linearly) isometric and isomorphic
Hilbert spaces.

	Beginning from now on in the present investigation, \emph{only
compatible Hilbert bundles will be employed}. For brevity, we shall call them
simply Hilbert bundles.

	Defining the \emph{Hermitian conjugate} mapping
\( \mor{A}_{x}^{\ddag}\colon \Hil\to \bHil_x \) of a mapping
\( \mor{A}_x\colon  \bHil_x \to \Hil \) by
	\begin{equation}	\label{2.3}
\langle \mor{A}_{x}^{\ddag}\varphi|\chi_x \rangle_x :=
\clangle \varphi | \mor{A}_x\chi_x\crangle,
\qquad \varphi\in\Hil,\quad \chi_x\in \bHil_x,
	\end{equation}
we find (see~(\ref{2.1}), the dagger denotes Hermitian conjugation in $\Hil$)
	\begin{equation}	\label{2.4}
\mor{A}_{x}^{\ddag} =
l_{x}^{-1}\circ \left( \mor{A}_x\circ l_{x}^{-1} \right)^\dag .
	\end{equation}

	We call a mapping $\mor{A}_x\colon\bHil_x\to\Hil$ \emph{unitary} if
it has an inverse $\mor{A}^{-1}$ and
	\begin{equation}	\label{2.5}
\mor{A}_{x}^{\ddag} = \mor{A}_{x}^{-1} .
	\end{equation}
Evidently, the isometric isomorphisms $l_x\colon\bHil_x\to\Hil$ are unitary
in this sense:
	\begin{equation}	\label{2.6}
l_{x}^{\ddag} = l_{x}^{-1}.
	\end{equation}

	The \emph{Hermitian conjugate}
mapping to a mapping $\mor{A}_{x\to y}$ in a set
\(
\{\mor{C}_{x\to y}\colon\bHil_x\to\bHil_y,\ x,y\in\base\}
\)
of mappings between the fibres of the bundle is a mapping
\( \mor{A}_{x\to y}^{\ddag}\colon \bHil_x\to \bHil_y \)
defined via%
	\begin{equation}	\label{2.7}
\langle \mor{A}_{x\to y}^{\ddag} \mor{\Phi}_x| \mor{\Psi}_y \rangle_y :=
\langle \mor{\Phi}_x| \mor{A}_{y\to x}\mor{\Psi}_y \rangle_x,
\qquad \mor{\Phi}_x\in \bHil_x,\quad \mor{\Psi}_y\in \bHil_y.
	\end{equation}
Its explicit form is
	\begin{equation}	\label{2.8}
\mor{A}_{x\to y}^{\ddag} =
l_{y}^{-1}\circ
\left( l_x\circ \mor{A}_{y\to x}\circ l_{y}^{-1} \right)^\dag
\circ l_x.
	\end{equation}
As $(\ope{A}^\dag)^\dag\equiv \ope{A}$ for any
$\ope{A}\colon \Hil\to\Hil$, we have
	\begin{equation}	\label{2.9}
\bigl( \mor{A}_{x\to y}^{\ddag} \bigr)^\ddag = \mor{A}_{x\to y} .
	\end{equation}

	 If
\(
\mor{B}_{x\to y}\in\{\mor{C}_{x\to y}\colon\bHil_x\to\bHil_y,\ x,y\in \base\}
\),
then a simple verification shows
	\begin{equation}	\label{2.10}
\left(\mor{B}_{y\to z}\circ \mor{A}_{x\to y} \right)^\ddag =
\mor{A}_{y\to z}^\ddag\circ\mor{B}_{x\to y}^\ddag, \qquad x,y,z\in \base.
	\end{equation}

	A mapping $\mor{A}_{x\to y}$ is called \emph{Hermitian} if
	\begin{equation}	\label{2.11}
\mor{A}_{x\to y}^\ddag = \mor{A}_{x\to y}.
	\end{equation}
A calculation proves that the mappings
\( l_{x\to y}:=l_y^{-1}\circ l_x \)
are Hermitian,
    \begin{equation}    \label{2.12-1}
( l_{x\to y} )^\ddag = l_{x\to y} .
    \end{equation}

	A mapping
\(
\mor{A}_{x\to y}\colon \bHil_x\to\bHil_y
\)
is called \emph{unitary}
if it has a left inverse mapping and
	\begin{equation}	\label{2.12}
\mor{A}_{x\to y}^\ddag = \mor{A}_{y\to x}^{-1},
	\end{equation}
where
\(
\mor{A}_{x\to y}^{-1}\colon \bHil_y\to\bHil_x
\)
is the \emph{left} inverse of
\(\mor{A}_{x\to y}\),
i.e.\
\(
\mor{A}_{x\to y}^{-1}\circ \mor{A}_{x\to y} := \id_{\bHil_x}
\).

	A simple verification by means of~\eref{2.7} shows the equivalence
of~\eref{2.12} with
	\begin{equation}	\tag{\ref{2.12}$^\prime$}	\label{2.12'}
\langle\mor{A}_{y\to x}\,\cdot\, | \mor{A}_{y\to x}\,\cdot\,\rangle _x
   = \langle\,\cdot\, | \,\cdot\,\rangle _y
   \colon\bHil_y\times\bHil_y\to\mathbb{C},
	\end{equation}
\ie the unitary mappings are fibre-metric compatible in a sense that they
preserve the fibre scalar (inner) product. Such mappings will be called
\emph{fibre\ndash isometric} or simply \emph{isometric}.

	It is almost evident that the mappings
	\begin{equation}	\label{2.13}
l_{x\to y}:=l_{y}^{-1}\circ l_x  \colon\proj^{-1}(x)\to\proj^{-1}(y)
	\end{equation}
are unitary, that is we have:%
\footnote{ %
The Hermiticity and at the same time unitarity of $l_{x\to y}$ is not
incidental as they define a (flat) linear transport (along paths or along the
identity mapping of $\base$) in $\HilB$. For details,
see~\cite{bp-BQM-1,bp-BQM-full}.%
}
	\begin{equation}	\label{2.14h}
l_{x\to y}^{\ddag}=l_{x\to y}=l_{y\to x}^{-1} .
	\end{equation}

	Let $\mor{A}$ be a morphism over $\base$ of $\HilB$, i.e.\
$\mor{A}\colon \bHil\to \bHil$ and $\proj\circ \mor{A} = \proj$,
and $\mor{A}_x:=\left.\mor{A}\right|_{\bHil_x}$.
The \emph{Hermitian conjugate}
bundle morphism $\mor{A}^\ddag$ to $\mor{A}$ is defined by (cf.~(\ref{2.7}))
	\begin{equation}	\label{2.15}
\langle \mor{A}^\ddag\mor{\Phi}_x | \mor{\Psi}_x \rangle_x :=
\langle \mor{\Phi}_x | \mor{A}\mor{\Psi}_x \rangle_x,
\qquad \mor{\Phi}_x,\mor{\Psi}_x\in \bHil_x.
	\end{equation}
Thus (cf.~(\ref{2.8}))
	\begin{equation}	\label{2.16}
\mor{A}_{x}^{\ddag} := \mor{A}^\ddag\big|_{\bHil_x} =
l_{x}^{-1}\circ\left(l_x\circ \mor{A}_x\circ l_{x}^{-1}\right)^\dag\circ l_x.
	\end{equation}

	A bundle morphism $\mor{A}$ is called
\emph{Hermitian}
if \(\mor{A}_{x}^{\ddag}=\mor{A}_x\) for every $\ x\in \base$, i.e.\  if
	\begin{equation}	\label{2.17}
\mor{A}^\ddag=\mor{A},
	\end{equation}
and it is called
\emph{unitary}
if \(\mor{A}_{x}^{\ddag}=\mor{A}_x^{-1}\) for every $\ x\in \base$, i.e.\  if
	\begin{equation}	\label{2.18}
\mor{A}^\ddag=\mor{A}^{-1}.
	\end{equation}
Using~\eref{2.15}, we can establish the equivalence of~\eref{2.17} and
	\begin{equation}	\label{2.19}
\langle\mor{A}\,\cdot\, | \mor{A}\,\cdot\,\rangle_x = \langle\,\cdot\, | \,\cdot\,\rangle_x
   \colon\bHil_x\times\bHil_x\to\mathbb{C}.
	\end{equation}
Consequently the unitary morphisms are fibre-metric compatible, \ie they are
\emph{isometric} in a sense that they preserve the fibre Hermitian scalar
(inner) product.

	To deal with the differentiable
properties of the employed Hilbert bundle $\HilB$, we
will require the  bundle space $\bHil$ and the base space $M$ be of
class $C^1$. Moreover, at the present level of development of quantum field
theory, we can  identify the base $\base$ with the 4\ndash dimensional
Minkowski space\ndash time. We shall require also the point\ndash trivializing
isomorphisms $l_x$ to have a $C^1$ dependence on $x\in\base$, \ie the mapping
$l\colon\bHil\to\Hil$ given by $l\colon u\mapsto l_{\proj(u)}u$ for
$u\in\bHil$, to be of class $C^1$ as a mapping between manifolds. A Hilbert
bundle with the last property will be called  $C^1$ bundle (or bundle of
class $C^1$).

\section{The bundle transport}
	\label{Subsect2.2}

	Suppose $(E,\pi,B,\mathcal{E})$ is a \field-vector bundle,
$\field=\field[R],\field[C]$, such that $E$, $B$, and $\mathcal{E}$ are
$C^1$ manifolds and its point\ndash trivializing isomorphisms $l_x$, $x\in
B$, are of class $C^1$, \ie it is of class $C^1$.

	\begin{Defn}	\label{Defn2.2}
The \emph{bundle transport} in a bundle $(E,\pi,B,\mathcal{E})$ is a mapping
$l\colon(x,y)\mapsto l_{x\to y}$, $x,y\in B$, where the mapping
	\begin{equation}	\label{2.20}
l_{x\to y}\colon \pi^{-1}(x)\to\pi^{-1}(y),
	\end{equation}
called \emph{(bundle) transport from $x$ to} $y$, is defined by
	\begin{equation}	\label{2.21}
l_{x\to y} := l_y^{-1} \circ l_x
	\end{equation}
with  $l_x\colon\pi^{-1}(x)\to\mathcal{E}$, $x\in B$, being the
point-trivializing isomorphisms of $(E,\pi,B,\mathcal{E})$.
	\end{Defn}

	Here are some frequently used properties of
the bundle transport
which are consequences of~\eref{2.21} and the linearity of
the isomorphisms $l_x$, $x\in B$:
	\begin{alignat}{2}				\label{2.22}
l_{x\to y}\circ l_{z\to x} &=
			l_{z\to y},&\qquad  x,y,z&\in B,
\\							\label{2.23}
l_{x\to x} &= \id_{\pi^{-1}(x)}, & x&\in B,
\\							\label{2.24}
l_{x\to y}(\lambda u + \mu v)
  &= \lambda l_{x\to y}u + \mu l_{x\to y}v,
	& \lambda,\mu &\in \field,\quad u,v\in{\pi^{-1}(x)},
\\							\label{2.25}
\left(l_{x\to y}\right)^{-1} &= l_{y\to x}.
	& x,y&\in B.
	\end{alignat}

	The bundle transport is Hermitian and unitary in a sense that such are
 $l_{x\to y}$, $x,y\in B$.

	A section $\mor{X}\in\Sec(E,\pi,B,\mathcal{E})$ is called
$l$-\emph{transported} (or ($l$-)constant) if for some (and hence any) $x\in
B$ and every $y\in B$ is fulfilled

       \begin{equation}		\label{2.26}
\mor{X}(y) = l_{x\to y} ( \mor{X}(x) )
	\end{equation}
where $\mor{X}(x):=\mor{X}_x$ with
$\mor{X}\colon x\mapsto \mor{X}_x$.
Such a section is uniquely defined by specifying its value at a
single point.  If $\ope{X}_0$ is a fixed  vector in the fibre
$\mathcal{E}$, the section $\mor{X}$ given via
	\begin{equation}	\label{2.27}
\mor{X}(x) = l_x^{-1} (\ope{X}_0)
	\end{equation}
is $l$-transported. Such sections will represent the states of quantum fields
in our approach.

	Let $(U,\kappa)$, $\kappa\colon U\to\field[R]^{\dim B}$, be a local
chart in a neighborhood $U$ of $x\in B$ and $x(\varepsilon,\mu)\in U$, where
$\varepsilon$ is a real number in a neighborhood of the zero and
$\mu=1,\dots,\dim B$, has coordinates
$\kappa^\nu(x(\varepsilon,\mu))=\kappa^\nu(x)+\varepsilon\delta_\mu^\nu$ with
 $\delta_\mu^\nu=1$ for
$\nu=\mu$ and $\delta_\mu^\nu=0$ for $\nu\not=\mu$. According to the general
formalism (see~\cite{bp-TM-general}
) in any $\kappa$ the bundle transport generates derivations
	\begin{equation}	\label{2.28}
\mor{D}_\mu\colon \Sec^1(E,\pi,B) \to \Sec^0(E,\pi,B)
\qquad \mu=1,\dots,\dim B
	\end{equation}
defined via their action on a $C^1$ section $\mor{Y}$ by
	\begin{equation}	\label{2.29}
(\mor{D}_\mu \mor{Y})(x)
:=
\mor{D}_\mu\big|_x (\mor{Y})
:=
\lim_{\varepsilon\to 0}
\frac {l_{x(\varepsilon,\mu)\to x}
\bigl(\mor{Y}(x(\varepsilon,\mu))\bigr) - \mor{Y}(x)}
      {\varepsilon} .
	\end{equation}

	The bundle transport also generates the mappings (derivations)	\begin{multline}	\label{2.30}
\Hat{\mor{D}}_\mu\colon
\bigl\{
\hmor{A}\in\MorfSec^1(E,\pi,B) \text{ is of class } C^1 \text{ and }
		\ \hmor{A}(\,\cdot\,)=\mor{A}\circ(\,\cdot\,)
        	\text{ with } \mor{A}\in\Morf_B^1(E,\pi,B)
\bigr\}
\\
\to
\bigl\{
\hmor{A}\in\MorfSec^0(E,\pi,B) \text{ is of class } C^0 \text{ and }
		\hmor{A}(\,\cdot\,)=\mor{A}\circ(\,\cdot\,)
		\text{ with } \mor{A}\in\Morf_B^0(E,\pi,B)
\bigr\} ,
	\end{multline}

	\begin{equation}	\label{2.31}
\Hat{\mor{D}}_\mu (\hmor{A}) := [\mor{D}_\mu,\hmor{A}]_{\_}
=
\mor{D}_\mu\circ \hmor{A} - \hmor{A}\circ \mor{D}_\mu
	\end{equation}
for any $C^1$ morphism $\hmor{A}$ of
$\Sec^1(E,\pi,B)$ generated by a morphisms $\mor{A}\in\Morf_B^1(E,\pi,B)$.

	The mappings $\mor{D}_\mu$ and $\Hat{\mor{D}}_\mu$ are derivations in a sense
that they are \emph{linear} and
	\begin{align}					\label{2.32}
\mor{D}_\mu(f\mor{Y}) &= \frac{\pd f}{\pd x^\mu} \mor{Y}
+ f \mor{D}_\mu(\mor{Y}),
\\							\label{2.33}
\Hat{\mor{D}}_\mu(\hmor{A}\circ \hmor{C})
&=
(\Hat{\mor{D}}_\mu(\hmor{A}))\circ \hmor{C} +\hmor{A}\circ(\Hat{\mor{D}}_\mu(\hmor{C}))
	\end{align}
where $f$ is a $C^1$ function on $B$ and $\hmor{C}$ and $\hmor{A}$ are $C^1$
morphisms of $\Sec^1(E,\pi,B)$.
	\begin{Prop}	\label{Prop2.1}
If $\mor{Y}$ is a $C^1$ section, $\mor{A}$ a $C^1$ morphism of $(E,\pi,B)$,
and $\hmor{A}(\,\cdot\,)=\mor{A}\circ(\,\cdot\,)$ is the morphisms of $\Sec(E,\pi,B)$
generated by $\mor{A}$, then
	\begin{align}					\label{2.34}
(\mor{D}_\mu \mor{Y})(x)
    & = l_x^{-1}\Bigl( \frac{\pd (l_x(\mor{Y}(x)))}{\pd x^\mu} \Bigr)
      =\bigl( l_x^{-1}\circ\frac{\pd}{\pd x^\mu}\circ l_x \bigr) (\mor{Y}(x)) ,
\\							\label{2.35}
\bigl( (\Hat{\mor{D}}_\mu \hmor{A})(\mor{Y}) \bigr) (x)
	&= \Big( l_x^{-1}\circ
		\frac{\pd (l_x\circ \mor{A}_x \circ l_x^{-1})}{\pd x^\mu}
	   \circ l_x \Bigr) (\mor{Y}(x)).
	\end{align}
	\end{Prop}

	\begin{Proof}
First of all, note that since $l_x(\mor{Y}(x))$ is a vector in the standard
fibre $\mathcal{E}$ and
\(
l_x\circ \mor{A}\circ l_x^{-1}
= l_x\circ \mor{A}|_{\pi^{-1}(x)}\circ l_x^{-1}
\)
is an operator in it, the partial derivatives of them are well defined. The
equality~\eref{2.34} is a simple consequence of~\eref{2.29} and~\eref{2.21}.
To verify~\eref{2.35} one should apply~\eref{2.31} to a  $C^1$ section $\mor{Y}$,
evaluate the result at $x$, and next to use~\eref{2.34}.
	\end{Proof}

	If $\{x^\mu\}\mapsto\{x^{\prime\,\mu}\}$ is a change of the
local coordinates, then  $\mor{D}_\mu$ and $\Hat{\mor{D}}_\mu$ behave like
basic (tangent) vector fields over $B$:
	\begin{equation}	\label{2.38}
\mor{D}_\mu\mapsto \mor{D}'_\mu
= \frac{\pd x^\nu}{\pd x^{\prime\mu}} \mor{D}_\nu,
\quad
\Hat{\mor{D}}_\mu\mapsto \Hat{\mor{D}}'_\mu
= \frac{\pd x^\nu}{\pd x^{\prime\mu}} \Hat{\mor{D}}_\nu .
	\end{equation}
This result, which follows from proposition~\ref{Prop2.1},
shows that one can define invariant `covariant derivatives' $\mor{D}_V:=V^\mu
\mor{D}_\mu$ and $\Hat{\mor{D}}_V:=V^\mu\Hat{\mor{D}}_\mu$ along a vector field
$V=V^\mu\frac{\pd}{\pd x^\mu}$ tangent to $B$.

	The bundle transport $l$ induces a natural `\emph{transport along the
identity mapping of} $B$' in the bundle of point\ndash restricted
morphisms $\morf_B(E,\pi,B)=(E_0,\pi_0,B)$. By this term we denote a mapping
$\lcl\colon(x,y)\mapsto \lcl_{x\to y}$, $x,y\in B$, where the mapping
\(
\lcl_{x\to y} \colon\pi_0^{-1}(x)\to\pi_0^{-1}(y) ,
\)
called \emph{transport from  $x$ to $y$ in} $\morf_B(E,\pi,B)$, is defined by
	\begin{equation}	\label{2.39}
\lcl_{x\to y} (\chi_x)
:=
l_{x\to y}\circ\chi_x \circ l_{y\to x}
=
l_y^{-1}\circ(l_x\circ\chi_x \circ l_x^{-1})\circ l_y
\colon\pi^{-1}(y)\to\pi^{-1}(y)
	\end{equation}
where $\chi_x\colon\pi^{-1}(x)\to\pi^{-1}(x)$. One can check that the
transports $\lcl_{x\to y}$ are \field\ndash \emph{linear} and satisfy the
basic `transport equations' (cf.~\eref{2.22}--\eref{2.25}):

	\begin{alignat}{2}				\label{2.40}
\lcl_{x\to y}\circ \lcl_{z\to x} &=
			\lcl_{z\to y},&\qquad  x,y,z&\in B,
\\							\label{2.41}
\lcl_{x\to x} &= \id_{ \{ \pi^{-1}(x)\to\pi^{-1}(x) \} }, & x&\in B,
\\							\label{2.42}
\bigl( \lcl_{x\to y} \bigr)^{-1} &= \lcl_{y\to x},
	& x,y&\in B.
	\end{alignat}
We shall call $\lcl$ the \emph{transport associated to the bundle transport}
$l$ (in $\morf_B(E,\pi,B)$).

	A morphism $\mor{A}\in\Morf_B(E,\pi,B)$ is said to be
$\lcl$-transported (or ($\lcl$-) constant) if the restrictions
$\mor{A}_x:=\mor{A}|_{\pi^{-1}(x)}$ satisfy the equation
	\begin{equation}	\label{2.43}
\mor{A}_y
=\lcl_{x\to y} (\mor{A}_x)
=l_{x\to y} \circ \mor{A}_x\circ l_{y\to x},
\qquad x,y\in B.
	\end{equation}
One can construct such kind of morphisms in the following way.
Let us fix $x\in B$ and
$\chi_0\colon\pi^{-1}(x)\to\pi^{-1}(x)$. The morphism
$\lcl_x(\chi_0)\in\Morf_B(E,\pi,B)$ such that
	\begin{equation}	\label{2.44}
\lcl_x(\chi_0)|_{\pi^{-1}(y)}
:= \lcl_{x\to y}(\chi_0)
= l_{x\to y} \circ \chi_0 \circ l_{y\to x}
	\end{equation}
is $\lcl$-transported due to~\eref{2.40}. Analogously, for fixed $x$, any
morphism  $\mor{A}\in\Morf_B(E,\pi,B)$ defines an $\lcl$-transported
morphism, denoted by $\lcl_x(\mor{A})$ and such that
	\begin{equation}	\label{2.45}
\lcl_x(\mor{A})
:= \lcl_x(\mor{A}_x) \colon y\mapsto (\lcl_x\mor{A})_y
:=\lcl_{x\to y}\mor{A}_x
=l_{x\to y}\circ \mor{A}_x\circ l_{y\to x}.
	\end{equation}

	Similarly to the transport $l$ considered above, the transport
$\lcl$ induces derivations $\lcD_\mu$ on the set
$\Sec^1(\morf_B(E,\pi,B))$ according to
	\begin{equation}	\label{2.46}
(\lcD_\mu \mor{A})(x)
:=
\lim_{\varepsilon\to 0}
\frac {\lcl_{x(\varepsilon,\mu)\to x}
	\bigl(\mor{A}_{x(\varepsilon,\mu)}\bigr) - \mor{A}_x}
      {\varepsilon}
	\end{equation}
with $x(\varepsilon,\mu)$ defined above and
$\mor{A}\in\Sec^1(\morf_B(E,\pi,B))$. Applying~\eref{2.39}, we find
	\begin{equation}	\label{2.47}
(\lcD_\mu \mor{A})(x)
	= l_x^{-1}\circ
	\frac{\pd (l_x\circ \mor{A}(x)\circ l_x^{-1})}{\pd x^\mu}
	   \circ l_x.
	\end{equation}

	If
$\chi\colon\Morf_B(E,\pi,B)\to\Sec(\morf_B(E,\pi,B))$
is the bijection mentioned in Subsect.~\ref{Subsect2.1},
 $\mor{A}\in\Morf_B(E,\pi,B)$, and
$\hmor{A}(\,\cdot\,)=\mor{A}\circ(\,\cdot\,)$, then (see~\eref{2.35})
	\begin{equation}	\label{2.48}
\bigl( (\Hat{\mor{D}}_\mu(\hmor{A})) (\mor{Y}) \bigr) (x)
=
\bigl(\lcD_\mu(\chi(\mor{A}))\bigr)(\mor{Y}(x)) .
	\end{equation}

	If, for every morphism $\hmor{A}\in\MorfSec\HilB$, we define a mapping
$\breve{l}_x \hmor{A}$ from the set $\Sec^1\HilB$ to the set of
mappings from $\base$ into $\proj^{-1}(x)$ by
	\begin{gather}	\label{2.49}
( ( \breve{l}_x \hmor{A} ) (\mor{Y}) ) (y)
:=
(l_{y\to x}\circ\hmor{A}_y) (\mor{Y}(y))
-
(\hmor{A}_x\circ(l_{y\to x}) (\mor{Y}(y)),
\qquad
\intertext{with $y\in\base$ and  $\mor{Y}\in\Sec\HilB$, we have the relation}
			\label{2.50}
( (\Hat{\mor{D}}_\mu\hmor{A}) (\mor{Y}) ) (x)
=
\Bigl\{
\frac{\pd}{\pd y^\mu} [ ( (\breve{l}_x \hmor{A}) (\mor{Y}) )(y) ]
\Bigr\}\Big|_{y=x} .
	\end{gather}
To prove it, one should insert~\eref{2.21} into~\eref{2.49}, to apply to the
obtained equality the operator $\frac{\pd}{\pd y^\mu}$, then to put $y=x$ and
to use~\eref{2.35}.

	Now we shall present some expressions in (local) bases which may be a
little more familiar to the physicists.
	 Sums like
$a^ib_i$ should be understood as a sum over $i$ and/or integral over $i$ if
the index $i$ takes countable and/or uncountable values; for some more
details --- see~\cite{bp-BQM-2}.

	Let $\{e_i(x)\}$ be a basis in $\pi^{-1}(x)$, $\{f_i\}$ a basis  in
$\mathcal{F}$ with $\pd_\mu f_i = 0$, and  $\Mat{l}_x$ and $\Mat{l}(y,x)$ be
the matrices of respectively $l_x$ and $l_{x\to y}$ in them, \ie if
 $l_x(e_i(x)) =: \Sprindex[l_x]{i}{j} f_j$
and
 $l_{x\to y}(e_i(x)) =: \Sprindex[l]{i}{j}(y,x) e_i(y)$.
Then
 $\Mat{l}_x:=[\Sprindex[l_x]{i}{j}]$
and
 $\Mat{l}(y,x):=[\Sprindex[l]{i}{j}(y,x)] = \Mat{l}^{-1}(y) \Mat{l}(x)$
(see~\eref{2.21}.) Then, ,for $\mor{Y}=\mor{Y}^i e_i$
from~\eref{2.29}, we get (cf.~\cite{bp-TM-general})%
\footnote{~%
In this paper we assume the Einstein summation convention: over indices
repeated on different levels a summation over their whole range is implicitly
understood.%
}
	\begin{gather}	\label{2.b1}
(\mor{D}_\mu\mor{Y})(x)
=
\Bigl(
\frac{\pd\mor{Y}^i(x)}{\pd x^\mu} +
	\Sprindex[\Gamma]{j\mu}{i}(x) \mor{Y}^j(x)
\Bigr) e_i(x)
 \\
(\mor{D}_\mu e_j)(x) =: \Sprindex[\Gamma]{j\mu}{i}(x) e_i(x)
	\end{gather}
with
	\begin{equation}	\label{2.b3}
\Gamma_{\mu}(x)
:=
[ \Sprindex[\Gamma]{j\mu}{i}(x) ]
=
\frac{\pd\Mat{l}(y,x)}{\pd x^\mu}\Big|_{y=x}
=
\Mat{l}_x^{-1}
\frac{\pd\Mat{l}(x)}{\pd x^\mu}
	\end{equation}
beingare the matrices of the \emph{coefficients} (resp.\ \emph{components})
$\Sprindex[\Gamma]{j\mu}{i}$
of $l$ (resp.\ $\mor{D}_\mu$).

	Under the changes
 $e_i\mapsto e'_i=C_{i}^{j}e_j$ and $x^\mu\mapsto x^{\prime\,\mu}$,
$C=[C_{i}^{j}]$ being non\ndash degenerate matrix\ndash valued  $C^1$
function, the matrices $\Gamma_\mu$ transform
into~\cite[eq.~(4.8)]{bp-TM-general}
	\begin{equation}	\label{2.b4}
\Gamma'_{\mu}(x)
:=
\Bigl(
	C^{-1}(x) \Gamma_\nu(x) C(x)
      + C^{-1}(x)\frac{\pd C(x)}{\pd x^\nu}
\Bigr)
\frac{\pd x^\nu}{\pd x^{\prime\,\mu}}.
	\end{equation}

	Applying~\eref{2.31} to some $C^1$ section $\mor{A}$ and taking into
account~\eref{2.b1} and~\eref{2.b3}, we find the matrix of
$\Hat{\mor{D}}_\mu\hmor{A}$ in $\{e_i\}$ as
	\begin{equation}	\label{2.b5}
[ (\Hat{\mor{D}}_\mu\hmor{A})_{i}^{j}]|_x
=
\Bigl(
\frac{\pd\Mat{\mor{A}}}{\pd x^\mu}
+ [\Gamma_{\mu}(x),\Mat{\mor{A}}]_{\_}
\Bigr)\Big|_{x}
=
\Mat{l}_{x}^{-1} \frac{\pd\Mat{\ope{A}}(x)}{\pd x^\mu} \Mat{l}_x
	\end{equation}
with $\Mat{\mor{A}}$ being the matrix of $\mor{A}$ in $\{e_i\}$. Here the
second equality is a consequence of
$\Mat{\mor{A}}=\Mat{l}_x^{-1}\Mat{\ope{A}}\Mat{l}_x$
which is equivalent to
$\hmor{A}_x=l_x^{-1}\circ\ope{A}(x)\circ l_x$.



\section{On bundle formulation of quantum field theory}
\label{Sect3}

	A quantum field $\bs\varphi$ is a operator-valued vector distribution
(generalized function), $\bs\varphi=(\bs\varphi_1,\dots,\bs\varphi_n)$ with
$n\in\field[N]$ and $\bs\varphi_i$, $i=1,\dots,n$, being operator\ndash valued
distributions called components of $\bs\varphi$, such that
	\begin{equation}	\label{3.3}
\bs\varphi(f)
=
\int_\base \Id^4y\sum_{i}^{} \varphi_i(y) f^i(y)
=
\sum_i \bs\varphi_i(f^i),
\quad
\bs\varphi_i(g)
:=
\int_{\base} \Id^4y \varphi_i(y) g(y)
	\end{equation}
for a vector-valued test function $f=(f^1,\dots,f^n)$ and
$f^i,g\colon\base\to\field$ being test functions. In this (symbolic) equation
$\varphi_i(y)$ are the components of the non-smeared field
$\varphi=(\varphi_1,\dots,\varphi_n)$ which (as well as $\bs\varphi(f)$) are
operators acting on the state vectors in the system's (field's) Hilbert space
$\ope{F}$ of states. Here and below, we implicitly adopt the Heisenberg picture
of motion.

	The discussion
in~\cite[subsect.~4.2]{bp-BQM-1} can \emph{mutatis
mutandis} be repeated here with respect to the state vectors. In short, this
leads to the replacement of the system's Hilbert space $\ope{F}$ of states with
a Hilbert fibre bundle $\HilB$ (of states) with Minkowski spacetime $\base$ as
a base, projection $\proj\colon\bHil\to\base$, fibres $\proj^{-1}(x)$ with
$x\in\base$, and the system's ordinary Hilbert space $\ope{F}$ as a (standard,
typical) fibre. Let $\{l_x : x\in\base\}$ be a set of (linear) isomorphisms
$l_x\colon\bHil_x\to\Hil$. Given an observer $O_x$ at $x\in\base$, if a
system is characterized by a state vector $\ope{X}\in\ope{F}$, the vector
	\begin{equation}	\label{3.4}
\mor{X}(x) := l_x^{-1}(\ope{X})
	\end{equation}
should be considered as its state vector relative to $O_x$. Hereof the state
vector $\ope{X}$ is replace with a state section $\mor{X}\in\Sec\HilB$ such
that $\mor{X}\colon x\mapsto\mor{X}(x)=l_x^{-1}(\ope{X})$.

	Accepting the above bundle description of states, to the bundle
description of quantum fields and, in general, to any operator
$\ope{A}(x)\colon\ope{F}\to\ope{F}$ can be applied arguments similar to the
ones in~\cite[subsect.~3.1]{bp-BQM-2}. At a point
$x\in\base$, the bundle analogue $\mor{A}_x$ of such a mapping $\ope{A}(x)$
can be defined by requiring the transition $\ope{A}(x)\mapsto\mor{A}_x$ to
preserve the scalar products,%
\footnote{ %
Physically this means independence of the observed values of the dynamical
variables of the way we calculate them.%
}
\ie
	\begin{equation}	\label{3.5}
\clangle \ope{X}|\ope{A}(x)(\ope{Y}) \crangle
=
\langle \mor{X}(x)|\mor{A}_x(\mor{Y}(x)) \rangle _x,
	\end{equation}
where $\ope{X},\ope{Y}\in\ope{F}$, $x\in\base$,
$\mor{X}(x)=l_x^{-1}(\ope{X})$,
$\mor{Y}(x)=l_x^{-1}(\ope{Y})$,
and $\clangle \,\cdot\,|\,\cdot\, \crangle$ and $\langle \,\cdot\,|\,\cdot\, \rangle_x$
are the scalar products in the Hilbert spaces $\ope{F}$ and $\proj^{-1}(x)$,
respectively, which are connected by (see Subsect.~\ref{Subsect2.2}
or~\cite[eq.~(3.1)]{bp-BQM-1})
	\begin{gather}	\label{3.6}
\langle \,\cdot\,|\,\cdot\, \rangle _x=
\clangle l_{x}\,\cdot\,|l_{x}\,\cdot\, \crangle , \qquad x\in \base
\\
	\tag{\ref{3.6}$^\prime$}	\label{3.6'}
\clangle \,\cdot\,|\,\cdot\, \crangle =
\langle l_{x}^{-1}\,\cdot\,|l_{x}^{-1}\,\cdot\, \rangle _x, \qquad x\in \base.
	\end{gather}
In particular, if $\ope{A}(x)$ is an operator representing a dynamical
variable $\dyn{A}$, equation~\eref{3.5} implies that the observed (mean)
value of $\dyn{A}$ is independent of the way we compute (or describe) it.
Combining~\eref{3.5} and~\eref{3.6'}, we derive
	\begin{equation}	\label{3.7}
\mor{A}_x = l_x^{-1}\circ\ope{A}(x)\circ l_x .
	\end{equation}

	In this way, we see that to an \emph{operator}
$\ope{A}(x)\colon\ope{F}\to\ope{F}$ there corresponds a \emph{morphism}
	\begin{equation}	\label{3.8}
\mor{A}\in\Morf_{\base}\HilB
	\end{equation}
of the system's Hilbert bundle with $\mor{A}|_{\proj^{-1}(x)}=\mor{A}_x$
given via~\eref{3.7}.

	In particular, the non-smeared components $\varphi_i$ of a quantum
field $\bs\varphi$ change to $\mor{\Phi}_i\in\Morf_{\base}\HilB$ with
	\begin{equation}	\label{3.9}
\mor{\Phi}_i|_{\proj^{-1}(x)} := l_x^{-1}\circ\varphi_i(x)\circ l_x.
	\end{equation}
Similarly, the smeared field components $\bs\varphi_i$ should be replaced
with mappings
	\begin{equation}	\label{3.10}
\bs{\mor{\Phi}}_i\colon x\mapsto \bs{\mor{\Phi}}_i|_x
:=
l_x^{-1}\circ\bs\varphi_i(\,\cdot\,)\circ l_x.
	\end{equation}
as a result of which the smeared field $\bs{\mor{\Phi}}$ becomes a mapping
	\begin{equation}	\label{3.11}
\bs{\mor{\Phi}} \colon x\mapsto \bs{\mor{\Phi}}|_x
=
l_x^{-1}\circ\bs\varphi(\,\cdot\,)\circ l_x,
	\end{equation}
so that
	\begin{equation}	\label{3.12}
\bs{\mor{\Phi}}_x(f)
=
\int_\base\Id^4y \sum_{i}^{}
	\lcl_{y\to x} (\mor{\Phi}_i(y))f^i(y)
	\end{equation}
where equations~\eref{3.3} and~\eref{3.9} were used and the transport $\lcl$
in $\morf_{\base}\HilB$ is defined via~\eref{2.39}.

	However, the above-introduced morphisms, like $\mor{A}$, are not the
exact objects we need. Here are two reasons for this. On one hand, since in
the ordinary theory the operators like, $\ope{A}$, act on state vectors,
like $\ope{X}$, we should expect the bundle analogue $\hmor{A}$ of $\ope{A}$
to act on a state section $\mor{X}$ producing again some section
$\hmor{A}(\mor{X})\in\Sec\HilB$. On another hand, since in the module
$\Sec\HilB$ there is a natural scalar product $\langle \,\cdot\, |\,\cdot\, \rangle$
with values in the $\field[C]$\ndash valued functions (see
Subsect.~\ref{Subsect2.2}), the r.h.s.\ of~\eref{3.5} should be identified
with the value at $x\in\base$ of the scalar product
$\langle\mor{X}|\hmor{A}(\mor{Y}) \rangle$.  Combining these ideas, we
conclude that
	\begin{equation}	\label{3.13}
\hmor{A}(\mor{X}) = \mor{A}\circ \mor{X} .
	\end{equation}
Hereof, the object $\hmor{A}$ is the morphism in $\MorfSec\HilB$ generated by
the morphism $\mor{A}\in\Morf_{\base}\HilB$ whose restriction on $\proj^{-1}(x)$
is~\eref{3.7}.

	Assuming $\hmor{A}$ to be the `right' bundle analogue of $\ope{A}$, we
conclude that
	\begin{equation}	\label{3.14}
\hmor{A}(\mor{X})\colon x\mapsto \hmor{A}(\mor{X}) |_x
= (\mor{A}\circ\mor{X})(x)
= \mor{A}_x(\mor{X}(x))
= l_x^{-1}( \ope{A}(x)(\ope{X}) ) ,
	\end{equation}
\ie the image of $\ope{A}(x)(\ope{X})$ according to~\eref{3.3} is exactly the
value at $x$ of $\hmor{A}(\mor{X})$ and
	\begin{equation}	\label{3.15}
\clangle \ope{X}|\ope{A}(x)(\ope{Y}) \crangle
=
\langle \mor{X}|\hmor{A}(\mor{Y}) \rangle|_x
	\end{equation}
which expresses the invariance of the scalar products when we replace state
vectors with state sections. Since the scalar products define the observed
(expectation, mean) values of physically observable quantities, the
last equality expresses the independence of these values of the way we
calculate them.


\section {Conclusion}
\label{Conclusion}

	The present investigation can be regarded as a continuation of the
fibre bundle formulation of quantum physics begun
in~\cite{
	bp-BQM-1,bp-BQM-2,
	bp-BQM-3,bp-BQM-4,
	bp-BQM-5}.
Here we have applied a slightly different approach to the (canonical)
quantum field theory in Heisenberg picture. The basic idea is the standard
Hilbert space of states to be replaced with a Hilbert bundle with it as a
(typical) fibre, then all quantities of the ordinary quantum field theory are
mapped into their bundle analogues by means of the bundle transport, which
is an internal object of any particular (locally trivial) bundle, or other
mappings build from it.

	However, the realization of such a procedure is not unique. The main
reason being that the quantum field theory has not a unique generally
accepted formulation.




\addcontentsline{toc}{section}{References}
\bibliography{bozhopub,bozhoref}
\bibliographystyle{unsrt}
\addcontentsline{toc}{subsubsection}{This article ends at page}

\end{document}